# A Review of existing GDPR Solutions for Citizens and SMEs*
## – preprint –

Erik Klinger, Alex Wiesmaier, and Andreas Heinemann

Hochschule Darmstadt – University of Applied Sciences, Germany

**Abstract.** The GDPR grants data subjects certain rights, like the right to access their data from companies, but in practice multiple problems exist with exercising these rights such as unknown data holders or interpreting the received data. Small and medium enterprises on the other hand need to facilitate the obligations given by the GDPR, but often lack proper systems, staff and other resources to do so effectively. For the GDPR to be effective in practice, these problems need to be addressed. With the work at hand we provide an overview of existing software solutions for these problems (from an internet research), discuss to which degree they solve the various problems and what issues remain.

## 1 Introduction

The GDRP is in effect since 2018 and changed the privacy landscape by requiring businesses to obtain consent of data subjects before their personal information may be processed and stored. These data subjects are also granted certain rights such as the right to access, rectify and delete their personal data. Implementing the respective duties on business side, as well as other requirements, and being able to demonstrate compliance is a complex task and especially challenging if done manually and without the help of digital tools.

Even though respective solutions had some time to mature over the past few years, there are still Citizen Problems (CP) for exercising given data subject rights as well as SME Problems (SP) for implementing the GDPR:

CP1 Citizens have to know who holds their personal data before they can exercise their data subject rights [193].
CP2 Citizens have to make separate requests while navigating different processes for different businesses [200].

---

* This research work has been partly funded by the German Federal Ministry of Education and Research and the Hessian State Ministry for Higher Education, Research and the Arts within their joint support of the National Research Center for Applied Cyber-Security ATHENE.

CP3 Citizens have to understand the data received in different formats [254] in order to make educated privacy-related decisions [252] [192].

SP1 Small and Medium Enterprises (SMEs) often do not have their own IT-experts and required knowledge to provide effective GDPR services, as well as overall limited resources [240][197].

SP2 SMEs often do not have proper systems for easy and automated operations on the data of their costumers or members, and instead use manual processing of paper or spreadsheets. (Working Hypothesis)

With these problems at the center, an internet search is performed and the existing landscape of technical solutions, for citizens that want to exercise their rights as well as for SMEs that need to comply with the GDPR, is reviewed.

### 1.1 Related Work

To the best of our knowledge there currently are no comprehensive reviews about the state of solutions for citizens to exercise their granted rights or solutions for SMEs to achieve compliance with the GDPR and solutions.

Research from the perspective of citizens focus mainly on performing and analyzing Data Subject Requests (DSRs) and their results, analyzing privacy policies and suggesting privacy enhancing technologies. Sørum et al. [249], Wong et al. [254] and Kröger et al. [218] made Subject Access Requests (SARs) and Right to Data Portability (RtDP) requests to companies, analyzed the request process and the received data. They found that sending DSRs and receiving their responses usually happens via e-mail, that the returned data is heterogeneous and that citizens have trouble understanding the data responses [252] [192]. The most comprehensive list of transparency enhancing tools is provided by Spagnuelo et al. [245]. However, the presented tools are mostly theoretical concepts, obsolete, proof of concepts or not related to the GDPR.

Global privacy control is "a proposed specification designed to allow Internet users to notify businesses of their privacy preferences, such as whether or not they want their personal information to be sold or shared. It consists of a setting or extension in the user's browser or mobile device and acts as a mechanism that websites can use to indicate they support the specification" [95]. This helps citizens to configure their preferences for all websites in one central place in their browser.

High-level guides on needed actions [196] [210] [215] and specific implementations [213] for SMEs do exist, but an comprehensive overview about existing technological solutions for these actions is missing. Research about the problems SMEs face in implementing the GDPR[240][242][197] do exist, but do not focus on the general state of technical solutions. Closest to the work at hand comes the privacy tech vendor report [112] by the International Association of Privacy Professionals (IAPP) which is the only found extensive list of privacy solutions for SMEs. Major functionality categories like consent managers, data mapping and DSR are explained, however, details about how these functionalities are provided or how they work are not given. Ryan et al. [239] analyze the vendors presented in an earlier version of said tech vendor report, but with focus on the accountability principle of the GDPR. The authors find that the majority of the GDPR compliance tools fail this principle and that most solutions are limited in their scope and cannot be integrated by other tools or integrate other tools.

In [236] we present a literature review on privacy dashboards for citizens and GDPR services for small data holders that complements the paper at hand with an academic point of view.

### 1.2 Research Questions

The research questions which the work at hand aims to answer and therefore will guide the search, analysis and review of the solution landscape are as follows:

RQ1 What solutions exist for citizens to exercise their rights granted by the GDPR?
  RQ1.1 How can citizens find out who has data about them?
  RQ1.2 How can citizens make sense of their received data?
  RQ1.3 How can these solutions be found by citizens?
RQ2 What solutions exist for SMEs to comply with the GDPR?
  RQ2.1 Which Off-the-shelve solutions exist for SMEs?
  RQ2.2 How can solutions be found by SMEs?
RQ3 How can products, services and interfaces be used by citizens and SMEs?
  RQ3.1 What are similarities and differences?
  RQ3.2 What are advantages and disadvantages?
  RQ3.3 How is data provided to citizens?
  RQ3.4 What is the result from DSRs?
RQ4 To what degree are the problems stated in 1 resolved? Which problems are still unresolved?

### 1.3 Methodology and Structure

The search itself is discussed in section 2. Lacking an academic methodology specific for non-academic internet reviews, the methodology used in the work at hand is an adaption of Kitchenham's and Charter's "Guidelines for performing Systematic Literature Reviews in Software Engineering" [216]. The rough outline of their method is as follows:

1. Define necessary research questions.
2. Define search terms and resources to be searched.
3. Define the selection criteria for the inclusion and exclusion of studies.
4. Define the study selection strategy which will determine how the selection criteria will be applied.
5. Develop quality assessment checklists and procedures to assess the individual studies
6. Develop the data extraction strategy which will determine how the information of each study will be obtained

7. Develop the strategy for synthesizing the extracted data.

The defined research questions are in the previous subsection 1.2. The defined search terms, selection criteria and selection strategy are explained in section 2. Because this review does not analyze studies but provided functionalities of software products, quality assessment checklists are skipped. The forms used for data extraction can be found in section A. The data extraction and synthesis are manually performed by the authors. The synthesis of the extracted data for citizens is provided in section 3 in the form of data tables and qualitative explanations categorized by provided functionality. The synthesis of the extracted data for SMEs is provided in section 4. We close the work with an evaluation and conclusion in section 6.

## 2　Search

Simple search terms without search operators are used to imitate citizens and SME employees searching for specific problems. The only exception is the term "analyze my personal Facebook data -"Facebook insights"", as results without the minus operator are dominated by results for businesses analyzing the audience of their Facebook pages. Searches are refined and adapted until the results are relevant. Tables 3 and 4 present the final search terms that were used to find the solutions.

The criteria for inclusion and exclusions of solutions, lists and guides is as follows: For inclusion, at least one right granted to citizens or one obligation of SME as defined in the GDPR has to be addressed. Furthermore, at least one of the research questions given in subsection 1.2 has to be addressed. The language has to be English or German. Solutions that only exist theoretically, are archived ,or open source without commits in the past year, are excluded.

As Google is the most prominent search engine, all searches are conducted on google.com. Per default, each result page contains 20 items. If at least three items on a result page meet the inclusion criteria, the next result page is considered as well. Otherwise, the search using the current search term is ended. All items that meet the inclusion criteria and are a type of solution are saved on the `extraction list`. If the item is a list or a guide, it will be saved in a separate list. After all search terms are used, all solutions mentioned on the previously saved lists and guides are matched against the inclusion criteria and either saved on the `extraction list` or archived. After search completion, the extraction forms (found in section A) are used on all solutions on the `extraction list`. List items that do not meet the inclusion criteria, or meet the exclusion criteria are archived. The filled extraction forms of each solution are saved and used as basis for the analysis in sections 4 and 3.

Tables 3 and 4 show how many lists or guides (LoG) and solutions for citizens respectively SMEs were found for each search term. Search terms targeted to find solutions for finding accounts lead to a relatively high number of guides. Searches for GDPR data request tools lead to the most results of solutions for SMEs. Amongst these results is the review platform g2 which contains the large majority (100 out of 132) of the found SME solutions [20]. Templates and generators for DSRs are prominent in German language while even direct searches in English lead to only a few results.

## 3  Solutions for Citizens

After the application of the exclusion criteria and after removing duplicates, 70 solutions and 9 lists and 30 guides remain. Table 1 gives an overview of the existing solutions for citizens which help with exercising their rights granted by the GDPR.

Table 1. Overview of solutions for citizens

| Functionality | Nr. | Language en/de/both | Type | Open Source/ Proprietary | Free | Automation |
|---|---|---|---|---|---|---|
| Account finders | 6 | 6/-/- | Website:4, Software:1 | OS:1, P:5 | 5 | Full:1 Semi:5 No:0 |
| DSR | 47 | 2/-/1 | List:3 Template: 44 | OS: 1 P: 2 | 47 | Full:0 Semi:0 No: 47 |
| Consent management | 8 | 7/-/1 | Extension:3 Software:1 App:1 PIMS[1]: 3 | OS:2, P:6 | 7 | Full:3 Semi:1 No:1 |
| Data breach notification | 7 | 4/1/2 | Website: 7 | OS:1, P:6 | 2 | Full:4 Semi:0 No:2 |
| Data broker opt-out | 6 | 6/-/1 | Website: 5 Software:1 | OS:0, P:6 | 0 | Full:3 Semi:1 No:0 |
| Privacy policy analyzer | 4 | 3/-/1 | Website:3 Extension: 2 | OS:2.5, P:1 | 4 | Full:2 Semi:1 No:0 |
| Data analyzer | 4 | 4/-/- | Website:2 Software: 1 App:1 | OS:1, P:3 | 4 | Full:0 Semi:4 No:0 |

Most solutions are web-based and do not have to be downloaded in order to be used. 18 solutions could be identified as open source, 28 as proprietary and for the remaining this categorization is not applicable as, for example, it makes no sense to call an email-template for a Data Subject Access Request (DSAR) open source. Table 5 shows found lists that collect privacy tools or general guides about increasing privacy. They primarily present privacy-friendly (i.e. collecting no data) alternatives to services like Google and Facebook or tools like VPNs and many most found tools are not included on these sites. Tools for helping citizens exercise their rights or understanding received data from DSARs or privacy policies are, with a few exceptions, not included on these lists.

### 3.1 Finding Data Holders

Cleaning up old accounts reduces the amount of companies that can have a data breach and leak personal data. 30 guides were found which guide citizens through steps to find their accounts and provide some supporting tools. There is a consensus between the found articles and guides that there is no single solution to reliably find all online accounts. Finding accounts is a manual process and tools can only support this process. The search also did not find a solution that can find all accounts across all sectors like social media, utilities and various services like mail providers, entertainment services or learning platforms. It also is impossible to be sure that all accounts were found. Having an inventory of as many accounts as possible can only be achieved in a forward manner by adding each new account to this inventory. Many found articles explicitly state that every account should be added to a password manager for this purpose, with the added benefit of having secure passwords for all services. The following is the synthesis of presented actions for citizens to find their accounts, of all found articles shown in Table 6.

1. Check saved passwords in used browsers and password managers
2. Check Google, Facebook, Twitter, etc. for accounts linked to their services
3. Use websites that check social media and other websites for a given username. Examples are listed in Table 7
4. Check people search engines like Truthfinder [177].
5. Check data breach notification sites shown in Table 12 with known usernames or emails to find companies with leaked data that include these usernames and email-addresses.
6. Search personal emails for registration mails or other mails that hint for an existing account. Keywords that are often mentioned are "confirm your email", "registration", "welcome to", etc. Clean Email [23] and Mine [126] help with this process but as they need access to the email account this provides a privacy risk as they have access to all personal emails. Mine is specifically made for finding accounts and displays accounts in a list and allows deletion requests to each service.
7. Use search engines with current and old usernames and email addresses
8. Use sites in Table 8 focused on deleting old accounts, which list companies and how to request deletion of the account, and look for known companies.

There are still open problems with these presented actions: It is a manual and time-consuming process and citizens have to remember all their current and past used usernames. Using third party tools to help with this process introduces new privacy risks as personal data is given to more services. This method can only find accounts which the citizen made him- or herself. Which official bodies have a citizen's personal data will probably not be found as well as data collected by websites through web-tracking and data that was sold to third parties.

Third party tracking tools listed in Table 7 can be used to find trackers used on a website and finding out where the personal data goes to. Controllers are also required by law to inform their data subjects about agreements with joint controllers and citizens can check the data notices and terms of services to find these joint controllers- Both of these methods are impractical because of the sheer amount of used third-party controllers and trackers. A tool that follows the chain of third party controllers and trackers was not found.

## 3.2 Data Subject Requests

Help for making DSRs in German is mainly provided in the form of templates for the request itself which are listed in Table 9. The majority of these templates provide the request text in the form of a basic text, a word-document or a PDF. This text has to be copied and sometimes filled out with contact information of the desired company and information which authenticates the citizen. Some solutions [77] [51] [92][128] [98] allow the citizen to choose the type of request, the service and type in their authentication information and generate the texts for each service. With two exceptions, www.selbstauskunft.net [121] and www.dsgvo-auskunft.org [64], citizens have to send the requests themselves. In these two cases the request is sent via fax. It is not clear why the request is not sent via email.

Table 10 shows the amount of found templates and generators for each type of request. Most of them exist for access and deletion request. It is to note, that the GDPR does not require DSRs to be made in a certain form and templates are not required.

Websites shown in Table 8 keep lists of companies with instruction about the respective deletion process. The requests have to be made by the citizen, but these sites provide unified guidance through the different processes. ControlMyID [28] and Redact [237] help to delete information in a targeted way, for example by deleting specific messages from chat services.

Data brokers collect and sell personal information and provide search functionalities for this data. Citizens could request each individual data broker to delete their personal information themselves or pay a service listed in Table 10 to do it for them. All these solutions are paid subscription services which automatically and continuously delete personal data of their clients from these data brokers. DeleteMe has a DIY opt-out guide for their most requested data brokers [78] and Optery has a list of covered data brokers [222], which offers some free help and guidance for citizens.

## 3.3 Consent management

The found solutions for consent management shown in Table 11 can be divided into three categories. The first category consists of browser extensions which centralize control of website trackers like re:consent [160], SuperAgent [170] and PrivacyBee [141]. They let citizens configure and review their consent preferences for trackers in one central place, enforce these automatically on each visited website and may warn the user if a website does not honor these preferences.

In the second category are Privacy Dashboard for android [142] and Windows privacy dashboard [185] which can be downloaded onto the device to manage local consent preferences. The android privacy dashboard lets the user control permissions, e.g. for microphone usage, and windows dashboard for consent of telemetry.

Personal Information Management Systems (PIMS) make up the third category. These let citizens decide whom they share their data with, for which purposes and for how long and also allow retracting information. They also allow keeping track of shared information[201].

### 3.4 Data Breach Notification

Table 12 shows a list of found solutions for data breach notification. Citizens can use free tools shown in to find data breaches for given phone numbers or email addresses. Additionally, services offer citizens to subscribe to in order to become informed as soon as the given email, phone number or username is found in future data breaches. With the exception of haveibeenpwned, these services cost money Usually, these services also offer opting customers out of data brokers as well or other services like malware protection. Even though data holders are legally required to inform their users about data breaches, these services allow to be informed about data breaches that companies don't know about or have not yet announced but are already leaked on the internet. However, because of this legal obligation and because there are services that are free, it can be questioned whether 2-4 dollars per month should to be spent on such a service

### 3.5 Privacy Policy Analyzer

Both PrivacySpy [155] and Tosdr [79] shown in Table 13 provide a collection of analyzed terms of services and privacy policies of various companies. For each service, an aggregated list of relevant points regarding the citizens privacy is displayed and a grade for privacy friendliness is issued. This helps citizens to understand privacy policies and their implications. Both are crowdsourced and open-source projects and also offer browser extensions to display the summaries for the currently visited website.

Pribot [10] uses AI to analyze privacy policies and provides a visualization of the privacy policies terms in order to help understand the policy. Because of this, it can be used to understand policies not covered by PrivacySpy and Todsr. However, it was not sure whether this service is still maintained in the earlier months of research and the website cannot be reached as of 17th. December 2022.

### 3.6 Analyzing data

Two ready-to-be-used solutions for data analysis are Accountanalysis for Twitter [211] and Applymagicsauce for Facebook, Twitter and LinkedIn [12]. Accountanalysis shows the information like daily rhythm and volume of tweets, used hashtags and relations to other Twitter users. Applymagicsauce claims to provide "insights on your personality, intelligence, leadership, life satisfaction and more". A third tool for analyzing data received from Facebook is available on GitHub [229], but needs to be compiled before use. This tool ranks friends by amount of send messages, most used words in conversations and overall statistics about messages like the total number of messages and words and amount of sent messages by time period.

TransparencyVis [174] allows analyzing received data from discord, Facebook, Google, Instagram, LinkedIn, Netflix, TikTok and Twitter it will be mentioned here. It categorizes data by type (e.g. account, media, location), shows a visual timeline of events mentioned in the data and indicates the sensitivity of individual information points.

Other results are articles found articles listed in Table 14 that explain how citizens can analyze their data using tools like python. However, these require basic programming knowledge

All of these tools need the downloaded data from the respective tools and provide citizens with the necessary steps.

## 4 Solutions for SMEs

After the application of the exclusion criteria and after removing duplicates, 132 solutions and 13 lists remain. Table 2 gives an overview of the found solutions for SMEs. In the case of DSRs, companies often do not clearly state what is offered. Some only explicitly mention facilitating SARs, but later explain that all forms of DSRs are supported. Other companies indeed seem to only provide SARs. Companies generally do not clearly distinct between privacy policies and privacy notices and use these terms synonymously for the privacy notice which informs the data subject about their data practices.

All the information, about the functionalities provided by the solutions, that is discussed in this section was taken from the websites of the respective services.

The found solutions can be divided into 3 major categories and 1 special category. The first category includes solutions focused on compliance for websites with tools for consent management for cookies and trackers, and templates for privacy notices. Adzapier [9] for example also supports DSRs.

In the second category are Privacy as a Service (PaaS) which offer a combination of tools for all aspects of GDPR compliance for businesses. Generally, these solutions are provided in cloud based Software as a Service (SaaS) model and serve their client as web application or desktop app. These solutions remove the need for SMEs to operate their own infrastructure to run GDPR solutions. This helps SMEs with a limited number of employees that missing knowledge to run technical infrastructure. As trade-of, the PaaS provider has to be trusted and this also involves an additional third party that interacts with their costumers' data.

The third category consists of solutions focused on single parts of GDPR compliance, for example DSR management only.

The special category is PIMS, as they do not only help SMEs comply, but also offer citizens one central platform for their data management as well. Con-trolMyId [28], DigiMe [62], Concord [36] and Meeco [99] fall into this category. These services also provide a ready to use user interface for SMEs to handle DSRs.

Table 2. Overview of Solutions for SMEs

| Functionality | Nr. | Language en/de/both | Type | Open Source/ Proprietary | Free | Automation |
|---|---|---|---|---|---|---|
| Inventory | 38 | 33/1/4 | SaaS: 14 Web: 24 SW: 2 Addon 2 | OS:1, P:38 | 1 | ? |
| DSAR+DSR | 31+70 | 26+54/ 3+6/ 2+10 | SaaS: 8+19, Web: 3+14 SW:16+27 Addon: 2+4 | OS:0+8, P:31+70 | 2+8 | Full:0+25 Semi: No:1+0 |
| Consent management | 40 | 32/2/6 | SaaS:12 Web 28: Addon: 2 | OS: 4, P:40 | 6 | ? |
| Cookie Consent | 38 | 27/2/9 | SaaS: 7 Web 31: | OS: 2, P:38 | 10 | ? |
| Data breach handling | 47 | 33/2/5 | SaaS: 10 Web: 30 | OS: 1, P:7 | 1 | ? |
| Self Service Portals | 24 | 20/-/4 | SaaS: 5 Web: 19 | OS: 3, P:24 | 5 | ? |
| Privacy Notices | 44 | 26/10/8 | SaaS: 6 Web: 26 SW: 5 Addon: 5 | OS:3, P: 44 | 18 | ? |

### 4.1 Data Inventory

All five analyzed solutions in section 5 offer building and managing a data inventory and are similar. This provides a reference framework for all the found data inventory solutions listed in Table 15.

The first common step is to connect to different systems via Application Programming Interface (API) connectors for databases like Postgres or MySQL, data processing systems like SAP, Salesforce, as well as SaaS and cloud providers like Amazon AWS and Microsoft 365. As shown in section 5, the amount of offered integrations differs between the five solutions.

Some companies, for example Centrl [145] provide a way to include data of systems for which no pre-built connectors exist, and some companies like Osano [184] do not. It may or may not be possible to add data manually to these systems.

After the data systems are connected, the five analyzed solutions discover data on these systems, identify and categorize personal information and map it to individuals. The degree of automation for this process varies. OneTrust does large parts of this on its own with human supervision, while others provide humans with guidance to do this process themselves. Tools that discover and map data automatically, also keep the resulting data inventory up to date by either continuously or periodically, repeating this process.

Except for Transcend were it is unknown, the finished data inventory of the analyzed solutions in section 5 can be accessed via a search interface and a central dashboard that contains information about the connected data systems, data flow between these systems and other information about the data. The data processing activities may also be recorded for the Record of Processing Activities (RoPA) demanded by the GDPR.

### 4.2 Data Subject Requests

The examples in section 5 also allow to drive a framework with actions for processing DSRs. The full list of found solutions for DSR is shown in Table 16

All examples provide web-forms that can be put on the SMEs website and serve as intake interface for data subjects to make DSRs. In cases like OneTrust [232] and Securiti these pre-built web-forms are customizable. These requests are automatically put into the request management system. In some cases, like Osano [184], automatic intake via email-forms is also provided.. Manually adding DSRs into these systems is also possible. A dashboard for providing an overview of and managing requests is also included. Other examples of solutions that provide a management system for keeping track of received requests and guide through the necessary steps to fulfill these requests are Ecomply [70] and Enactia [73].

For the actual processing of DSRs, different levels of automation exist. In cases like Transcend [173] DSRs can be fulfilled fully automatic. In the case of Osano [184] SAR can only be partially automated and require human intervention, with the argument being that fully automated processes might make mistakes and process the DSR incorrectly. Thus, needing human intervention to ensure requests are fulfilled correctly.

### 4.3 Consent Management

Tools for consent management can be generally classified into systems for cookie consent only like Osano and Transcend, and systems for general consent like OneTrust and Securiti. The full list of found solutions for consent management is shown in Table 17 The examples give an outline for consent management solutions.

Osano, Transcend and Securiti offer to automatically scan the website for cookies and trackers to build an inventory. Identified trackers have to be classified by purpose of processing This is partly done automatically by matching them against databases containing existing tracking technologies and manually in cases where no match is found. This process is in some cases done continuously or by periodic repetition in order to keep the list of classified trackers up to date.

All examples captured consent via a cookie banner or a consent management widget on the website which may be auto-populated by the inventory of trackers. Depending on the solution, data subjects can give consent either accept-all/decline-all [145], by category (i.e. essential, analytics, functional, marketing) [149] or granular for each tracker. In all examples, these cookie banners are localized and available in different languages.

Clarip [144] and Centrl [145] emphasize, that consent from other sources than cookie banners can be captured. Given examples are forms for email opt-outs and "do not call" -lists from marketing automation as well as Customer Relationship Management (CRM) systems, and related tools in the case of Centrl and "IoT and third-party social media platforms" for Clarip. No further details were found. Both also emphasize their API which allows their consent management to be integrated into other consent management and capture tools as well.

Internally consent management is used to store a history of consent and preferences in order to be able to demonstrate that the SME has consent of the individual to process their data. It is also used to determine on which systems individual data subjects' data may or may not be collected, processed and stored. Dashboards can be used to gain a statistical overview about given consent. For example, the total opt-in/opt-out ratio, do not sell over time and preferences by category in the case of Centrl.

### 4.4  Data Notices

Solutions for data notices can be divided into templates and generator, and document management systems. The full list of found solutions is shown in Table 18 In some cases, companies did not further specify the provided functionalities and only mentioned providing some form of template or management system.

Templates for data notices generally provide a textual framework with necessary legal wording, containing gaps that have to be filled by the SME. The information that needs to be filled in are information about what the data notice is used for (e.g. a website or an app), which personal information is collected, for which purposes and to which regulations the notice has to comply to. Generators use a web-form or questionnaire which guides through this process and generate the data notice in the end. In both cases the generated policy can be downloaded or is sent via email and can be embedded on the SMEs website.

Some allow the generated policy to be used for free without offering legal insurance and offer this insurance in exchange for payment. Examples are TermsFeed [171], which can also generate Terms & conditions, cookie policies, terms of service and disclaimers. Cortina consult[53] is a German site offers a template. Seers [163] can also generate different types of policies.

Some policies have dynamic elements. CookiePro or Complianz[150] for example can update the cookie policy based on currently used cookies on the website and Datanships [42] claims that their privacy policies update automatically when laws are changed.

Document management systems for policies and notices provide one central place for creating, updating and deploying privacy notices and policies on websites. Examples for this are Dataships [42], Clym [149], Wrangu [33], Privacytools [151] and CookiePro [80]. CookiePro also is able to scan the SMEs websites in order to locate existing policies and build an inventory, which helps to identify websites with missing policies. Some solutions offer version management for updating the policies and notices and keeping track of pats changes. Clym also mentions that their policies can be hosted separately and then be embedded on the SMEs website. The link to the hosted policies can also be shared with partners which allows them to also have an up-to-date policy from the SME. This may be helpful for joint controllers. CookiePro allows the created policies to be integrated in content management systems with the use of plugins. Policies and notices can be deployed "across domains or apps via a tag manager or mobile SDK".

### 4.5 Data Subject Portals

The full list of found solutions for data subject portals is shown in Table 19 The detailed analysis in section 5 shows, offered data subject portals are used as central interaction point between the SME and its data subject for making DSRs, displaying data policies and managing consent preferences. Ethyca [38] is another example for this. Trunomi [41] also claims to display information about the users' data and DPOrganizer [94] claims to display information about " personal data processing" and "per visitor customized and layered information". Both do not give further explanations or details. In some cases [173][162], the portal is also used for communication during the processing of the DSR as well as transmitting and downloading the answer and associated data. All in all, companies only offer vague descriptions about exact functionalities and workings of their offered portals.

### 4.6 Data Breach Handling

The analysis of OneTrust and Securiti in section 5 provide two cases for data breach handling and give an outline of how the handling of data breaches is assisted by tools.

First, both of these tools offer an interface containing a form for reporting data breach incidents. In these cases, pre-built forms are offered for this. Wrangu [33] is another tool that also offers reporting this way.

OneTrust and Securiti both offer an overview dashboard, which lists incidents to keep track of them and the 72-hour deadline. Wrangu and RealCGR [158] also offer this. OneTrust, Securiti and Wrangu also offer a detailed view for each incident, which display necessary steps to process the incident. RealCGR may offer this also, but does not mention it on their website.

Both OneTrust and Securiti also offer templates for the breach notification. The full list of found solutions is shown in Table 20. Many companies remain vague about the actual functionalities and only claim to handle data breaches in an GDPR-compliant way.

## 5   Comparison of Five Solutions for SMEs

Due to the large number of products for SMEs and the fact that many of these products are only explained vaguely, five SME solutions are analyzed in more detail in this chapter. We look at how the provided functionalities are implemented by the respective products and how they can be used by the SME and their data subjects.

For the determination of these five solutions, the term „top 5 privacy management solutions" is used to find comparison and ranking websites. As there is no consensus about the top five solutions between these websites, the occurrence of each mentioned solution is counted and used as metric. The most often mentioned products are OneTrust, TrustArc, Osano, Transcend, Datagrail, and Securiti. TrustArc's public facing website does not contain detailed information about how their product works or how exactly it can be used. Because they also do not provide a public documentation which could contain further information, TrustArc is excluded from this analysis.

**The informational basis for this analysis are the public facing product sites and provided documentation. If not otherwise indicated, claims about products and pictures are taken from the respective websites.**

Table 21 and Table 22 show the comparison of the five solutions in tabular form. The five analyzed solutions are generally quite similar. Except for OneTrust, the analyzed companies did only provide general and basic information about the capabilities of their products and how they are set up and used. Osano, Transcend and DataGrail have their documentation publicly available on their website which provide detailed information. Even then, technical specifics are generally not given, except for Transcend which provide technical explanations and diagrams about their solutions. The lack of detailed information most likely is due to their emphasis on demos. Each company provides the opportunity to schedule a demo to get a better view of their product.

For building the data inventory, all offer pre-built integrations to connect to data systems and SaaS vendors in order to detect and classify contained personal information on these systems and build the data inventory. This process is mostly automatic, but may require additional help by humans, mostly for quality control of automated processes. An exact side by side comparison of provided integrations is not possible, as not all solutions show available solutions per category, e. g. data discovery and data mapping and consent management. The data inventory is kept up to date either through periodic or continuous repetition of discovery and classification. Notable differences are Osano, for which the ability to build custom integrations if no pre-built ones exist was not found, and Securiti where no information about building custom integrations was found.

For DSRs, overall the solutions are quite similar. All offer intake via webforms and automate fulfillment either fully or partially by using the integrations and interacting with the data inventory. Osano also offers intake via email and transcend also offers an API. For authentication, the minimum requirement for OneTrust, Transcend and Datagrail is the verification of the requesters-email. Other forms of authentication are supported, ranging from logging in, questions about the data and official identification documents. What exactly is used, is up to the SME. All DSR-types are supported. Securiti only mentions access and deletion request explicitly, but because of the breadth of their product catalog it may be assumed that they also support all types. None of the companies provided information about the data types and its representation sent for DSAR if data is found. The reason for that may be that this is up to the SME. It is most likely, that none of the businesses offer extraordinary ways of making data of SARs extra transparent and understandable to the data subject, for example by providing a visual diagrams, as such a capability would most likely be communicated on the websites of the company. Such information was not found on any website. Except for DataGrail which offers a download link via email, all companies use data portals for sending any data for DSRs to the data subject. There was no hint about the capability of importing data for portability requests on any website. Is it likely, that data can be added manually, but automatic means were not mentioned.

Data breach management is only offered by OneTrust and Securiti. Both provide intake forms, a management system which guides through necessary steps and provides an overview of deadlines and templates for the notification. What the notification the data subjects receive looks like is unknown.

Except for DataGrail, the companies provide management of consent for at least website trackers. OneTrust and Osano also provide managing consent on mobile apps. Also, OneTrust and Securiti offer to integrate and manage consent form other sources like consent for email newsletters using a separate form. With exception to OneTrust where it is unknown, a scan for existing cookies is provided and found trackers are matched against a database for automatic classification. Differences are that Osano offers only an initial scan while transcend and security offer periodic or continuous scans.

Templates for notices are provided by 3 of 5 solutions, but only OneTrust and security offer document management for data notices. They also offer dynamic population of these notices with cookie information and Securiti and transcend also offer the inclusion of information from the processing policies that where defined in the system.

## 6 Evaluation and Conclusion

### 6.1 Solutions for Citizens

In section 3 we give an overview of solutions for citizens by functionality for exercising their rights  (**RQ1**) thereby showing that tools for assisting citizens to exist, but still need to be improved.

Finding accounts (**SP1**, **RQ1.1**) is partially solved by a manual step-by-step process supported by tools, created by combining found articles listed in Table 6. However, it is not possible to find all accounts in retrospect. Remembering the accounts or used usernames and email addresses is still required. A tool for following the trail of data shared between companies to find data holders other than the one for the initial account is still missing.

The problem of having to make separate request while navigating different processes per data holder (**SP2**) is partially solved by DSR-generators that provide a list of available companies for which they can create and send requests. However, even the best version of these tools would require individual authentication to each company. One outlined solution for this is to integrate a DSR-tool into password managers to make requests to all saved accounts using their saved login-credentials.

For consent management, Global Privacy Control (GPC) [95] might be the solution for citizens to exert their consent preferences to all services in a unified way, if it is extended to cover not only consent on websites but also other sources like mobile apps.

The problem of understanding data received in DSRs (**CP3**, **RQ1.2**) is the least solved, as only a few tools help to analyze the data for the biggest services like Facebook, and a general purpose analyzing tool is missing as well as a solution for cross-referencing data from multiple sources.

Citizens can find tools (**RQ1.3**) via Google-search and found articles. Most found websites dedicated to collect and list privacy tools do only provide a couple of solutions for GDPR [152] [153] [154] [18] [6].

The concept of PIMS could help with all three of these problems, as it could be the central place for citizens to have a list of companies with access to their data, a unified way to manage consent preferences and make DSRs to individual or all companies. As central place for all data this would also allow analyzing available data and educate about the implications of sharing the data. Because of their potential, this concept should be researched further to point out existing problems and present solutions how to solve them.

With all the above, **RQ4** is answered for citizen solutions.

### 6.2 Solutions for SMEs

The work at hand provides a systematic overview of the 130 found solutions for SMEs listed in section D. The found solutions cover all necessary aspects of GDPR compliance for the interaction between SME and citizens, at least on the surface. Namely, building a data inventory, handling DSRs, managing consent, managing data notices and data breaches, and for data subject portals. Together with the comparison of five examples in section 5 to show how these tools work in a little more detail **RQ2** is answered.

Concerning existing off-the-shelve solutions for SMEs (**RQ2.1**), the tools analyzed in section 5 are usable examples. However, in these cases off-the-shelve does not mean that the SME will be automatically compliant after buying one of these solutions, but rather that tools are ready to be set up to interact with the SME's systems. SMEs still need to set up connectors to their data systems, built workflows for processing DSR and assign employees for the manual parts that are not automated.

The large amount of found solutions shows that SMEs can find GDPR solutions via Google and collection sites like G2 [20]. However, the work at hand does not judge on the completeness or quality of comparison lists and collection sites for GDPR solutions for SMEs and **RQ2.2** should be considered unanswered in this regard. Future research could examine the process of finding and choosing a suitable solution for SMEs, considering that the websites of analyzed examples in section 5 do not provide all details of their product.

The comparison in section 5 shows that provided capabilities of data inventory systems and DSR systems are similar, same for consent management with notable differences being the inclusion of consent from custom forms and mobile apps. In future, more solutions should be tried out and analyzed in more detail to include more solutions in the comparison.

**RQ3.2** is not answered in depth by the work at hand. In future, the analyzed solutions should be tried out in practice to properly compare advantages and disadvantages.

To answer **RQ3.3**, except for Datagrail, the data from DSAR is provided to data subjects via the privacy portal on their website. DataGrail sends a download link via email. Information about what the result is, is not provided (**RQ3.4**). Again, in future, the analyzed tools should be tried out in practice to collect more information about this.

As the examples in section 5 are provided as SaaS, SMEs do not need their own IT-experts to run these GDPR tools (**SP1**). However, as the examples are complex pieces of software and need to interact with internal data systems, employees might need to be trained in order to set these tools up correctly. The offered training by OneTrust [232], Osano [184] and Securiti [162] might help with this problem. Future research could examine costs and benefits of using these solutions in context of the overall limited resources of SMEs.

section 5 also shows that solutions for automated operations on personal data of customers and employees (**SP2**) do exist and manual processing using paper and spreadsheets is not necessary.

Processes for initial digitizing and importing the data in order to use these example solutions is not investigated in the work at hand. This should be examined in the future, as this is necessary for SMEs that currently process their data manually.

With all the above, **RQ4** is answered for SMEs.

## A   Extraction Forms

The used data extraction sheet for solutions is the following:

**Website** Link to the solution
**Search term + page** On which page using which search term was the solution found?
**Type** Is this solution SaaS, a plugin, add-on, regular software, a web application, etc.
**Functionalities** Which functionalities does the solution provide? E.g. Data discovery, DSR handling, consent management, etc.
**Other** Noteworthy data points. For example, if the solution provides an API for developers.
**Addressed problems** Which problems stated in the introduction are addressed by this solution?
**Addressed rights** Which data subject rights does this solution address?
**Open Source?** Is this solution open source or proprietary? If it is open source, is it still maintained?
**Cost** How much does this solution cost?
**Language** In which languages is this solution available?

The used data extraction sheet for lists and guides is the following:

**Website** Link to the list or guide
**Search term + page** On which page using which search term was the list or guide found?
**Presented Solutions** List of solutions this list or guide provides.
**Addressed problems** Which problems stated in the introduction are addressed by this?
**Addressed rights** Which data subject rights does this address?
**Language** In which languages is this written?

# B  Search

Table 3. English search terms

| Search Term | RQ Problem | User: LoG | User: Sol. | SME: LoG | SME: Sol. |
|---|---|---|---|---|---|
| Find my accounts | RQ1.1 | 13 | 2 | - | - |
| Finding online accounts | RQ1.1 | 13 | 1 | - | - |
| Find My Information Online | RQ1.1 | 3 | 4 | - | - |
| Website that knows your information | RQ1.1 | 2 | 1 | - | - |
| Account finder | RQ1.1 | 0 | 0 | - | - |
| Request templates | RQ1 | 1 | 2 | - | - |
| GDPR data request tools | RQ1, RQ2 | 0 | 1 | 8 | 19 |
| Privacy Tools | RQ1 | 10 | 1 | - | - |
| Privacy Dashboard | RQ1, RQ2 | 13 | 3 | - | - |
| privacy policy analyzer | RQ2, RQ1 | 0 | 2 | - | - |
| interactive privacy policy | RQ1, RQ2 | 0 | 0 | 1 | 2 |
| analyze my private data | RQ1.2 | 0 | 1 | 0 | 0 |
| Analyze my account data | RQ1.2 | 0 | 1 | - | - |
| Analyze my data | RQ1.2 | 0 | 0 | 0 | 0 |
| GDPR analyze my data | RQ1.2 | 0 | 0 | 0 | 0 |
| analyze my google data | RQ1.2 | 0 | 2 | - | - |
| analyze my youtube data | RQ1.2 | 0 | 3 | - | - |
| analyze my personal Facebook data - "Facebook insights" | RQ1.2 | 0 | 1 | - | - |
| analyze my personal amazon data | RQ1.2 | 0 | 0 | - | - |

Table 4. German search terms

| Search Term | RQ Problem | User: LoG | User: Sol. | SME: LoG | SME: Sol. |
|---|---|---|---|---|---|
| Meine Accounts finden | RQ1.1 | 5 | 0 | - | - |
| dsgvo auskunft anfordern | RQ1 | 0 | 14 | - | - |
| DSGVO Auskunft | RQ1 | 0 | 15 | - | - |
| DSGVO Anfrage Generator | RQ1 | 1 | 15 | - | - |
| Datenschutzerklärung | RQ2 | 0 | 6 | - | - |
| Datenschutzdashboard | RQ1, RQ2 | 0 | 2 | - | - |
| DSGVO DSAR Software | RQ1, RQ2 | - | - | 3 | 13 |
| "interaktive" datenschutzerklärung | RQ1, RQ2 | 0 | 0 | 0 | 0 |

# C   Solutions for Citizens

Table 5. Lists of privacy tools

| Name / Company | GDPR Tools | Ref. |
|---|---|---|
| PCMag | Optery, PrivacyBee, IDx, DeleteMe, Bitdefender | [2] |
| Aboalarm | JustDeleteMe, DeleteMe | [2] |
| awesome-humane-tech | ReConsent, Facebook Data Analyzer | [18] |
| choosetoencrypt | ToS;DR | [208] |
| PCWorld | SuperAgent | [6] |
| cogipas | JustDeleteMe, HaveIBeenPwned, ApplyMagic Source | [227] |
| MeAndMyShadow | tools deprecated | [102] |
| PrivacyGuides | ToS;DR | [152] |
| PrivacyTools | Redact | [153] |
| PrivacyToolsList | Sherlock, WhoTracksMe | [154] |

Table 6. Articles for citizens about finding accounts

| Name / Company | Ref. |
|---|---|
| **Finding Accounts** | |
| Aboalarm | [219] |
| supereasy | [203] |
| supereasy 2 | [202] |
| Avast | [60] |
| noypigeeks | [207] |
| Cybernews | [74] |
| Seon | [75] |
| FutureZone | [209] |
| Alphr | [105] |
| Helpdeskgeek | [106] |
| Avast 2 | [107] |
| Joincake | [109] |
| Passwordbits | [108] |
| Avast 3 | [111] |
| Getapp | [110] |
| Medium | [231] |
| PC Welt | [246] |
| Consumerreports | [143] |
| Knowrechie | [238] |
| Whoat Where Why | [251] |
| Techbook | [182] |
| Makeuseof | [253] |
| Verbraucherfenster | [178] |
| techboomers | [76] |

Table 7. Solutions for finding accounts

| Name / Company | Functionality | Price | Ref. |
|---|---|---|---|
| Checkusernames | Check username | Free | [21] |
| Knowem | Check username | Free | [120] |
| Namechk | Check username | Free | [129] |
| Truthfinder | People Search | Free | [177] |
| Sherlock | Find social media account | Free | [165] |
| Clean Email | Find registration mails | from 35.69€/year | [23] |
| Mine | Scans mails | Free? | [126] |
| Privacy.net | Identify web-trackers | Free | [247] |
| ProPrivacy | Identify web-trackers | Free | [172] |
| WhoTracksMe | Identify web-trackers | Free | [181] |

Table 8. Solutions for data subject requests

| Name / Company | Functionality | Price | Notes | Source |
|---|---|---|---|---|
| Accountkiller | List of deletion instructions | Free | | [100] |
| JustDeleteMe | List of deletion instructions | Free | | [118] |
| SimpleOptOut | List of instructions to restrict processing | Free | | [59] |
| BreachGuard | Data Broker Opt-Out | 43.00€/year | | [114] |
| DeleteMe | Data Broker Opt-Out | $129/year | | [161] |
| IDX | Data Broker Opt-Out | $129.92/year | | [26] |
| Incogni | Data Broker Opt-Out | 59.48€/year | | [117] |
| Optery | Data Broker Opt-Out | $99+/year | Free self-service removal tool | [223] |
| PrivacyBee | Data Broker Opt-Out | $197/year | | [141] |

GDPR Solutions Review   23

Table 9. Templates and generators for making data subject requests

| Name / Company | Type | Ref. |
|---|---|---|
| **Templates** | | |
| Anwalt Article | SAR | [189] |
| Arbeiterkammern Article | SAR, Deletion | [50] |
| Datenschutz.org Article | Deletion | [13] |
| DPA-Baden-Württemberg | SAR | [228] |
| Datenschutzexperte | SAR | [206] |
| DPA Hessen | SAR | [15] |
| Datenschutz.org | SAR | [16] |
| DPA Austria | All | [63] |
| Heise.de | SAR, Deletion | [233] |
| Hensche.de | SAR | [212] |
| wbd | SAR | [244] |
| Verbraucherzentrale | All | [93][230][115] |
| wko | SAR | [191] |
| **Generators** | | |
| auskunftsbegehren.at | SAR | [1] |
| Datenanfragen | SAR | [51] |
| Datenschmutz | SAR | [92] |
| Frag den Dienst | SAR | [77] |
| Hdgdldsgvo | SAR, Deletion, Restriction | [98] |
| patientenrechte-datenschutz.de | SAR | [122] |
| MyDataDoneRight | SAR, Deletion, Correct, Portability | [128] |
| **Generators that also send the DSR via fax** | | |
| DSGVO-Auskunft | SAR | [64] |
| Selbstauskunft | SAR | [121] |

Table 10. Supported request types by templates and generators

| request type | Templates | Generators |
|---|---|---|
| Access | 14 | 9 |
| Deletion | 7 | 3 |
| Restriction | 3 | 1 |
| Rectification | 3 | 0 |
| Objection | 3 | 0 |
| Portability | 1 | 0 |

Table 11. Solutions for consent management

| Name / Company | Functionality | Price | Source |
|---|---|---|---|
| BitsAboutMe | PIMS, Data Analysis | Free | [169] |
| Control.My.ID | PIMS, Deletion | Free | [28] |
| Digi.Me | PIMS | Free | [62] |
| Meeco | PIMS | Free | [99] |
| Redact | Delete posts on services | Free | [237] |
| PrivacyBee | Unified Consent Control | $197/year | [141] |
| re:consent | Unified Consent Control | Free | [160] |
| Super Agent | unified Consent Control | Free | [170] |
| Windows privacy dashboard | Consent Control | Free | [185] |
| (Android) Privacy Dashboard | App Permission Control | Free | [142] |

Table 12. Solutions for data breach notifications

| Name / Company | Functionality | Price | Notes | Source |
|---|---|---|---|---|
| cybernews tool | Breach Check | Free | | [135] |
| Avast Hackcheck | Breach Check | Free | Automatic if paid | [61] |
| Have I Been Pwned | Automatic Notification | Free | | [97] |
| Bitdefender | Automatic Notification | 69.99€/year | | [169] |
| BreachGuard | Automatic Notification | 43.00€/year | | [114] |
| IDX | Automatic Notification | $129.92/year | | [26] |
| PrivacyBee | Automatic Notification | $197/year | | [141] |

Table 13. Solutions for understanding privacy policies

| Name / Company | Functionality | Price | Notes | Source |
|---|---|---|---|---|
| Privacy Spy | List of Summaries | Free | +Extension | [155] |
| ToS;DR | List of Summaries | Free | +Extension | [79] |

Table 14. Solutions for analyzing data from data subject requests

| Name / Company | Functionality | Which data | Price | Open Source? | Source |
|---|---|---|---|---|---|
| ApplyMagicSauce | Analyzer | Facebook, Twitter, LinkedIn | Free | no | [12] |
| Luca Hammer | Analyzer | Twitter | Free to 15€/month | no | [211] |
| facebook_data_analyzer | Analyzer | Facebook | Free | need to be compiled | [229] |
| TransparencyVis | Analyzer | Facebook, Discord, Google, Instagram, LinkedIn, Netflix, TikTok, Twitter | Free | No | [174] |
| Bart Wronski Article | Guide | Google | Free | - | [194] |
| TowardsDataScience | Guide | Youtube | Free | - | [199] |
| DataQuest | Guide | Facebook | Free | - | [204] |
| TowardsDataScience 1 | Guide | Google | Free | - | [198] |
| DataGoblins | Guide | Youtube | Free | - | [104] |
| jovian.ai | Guide | Youtube | Free | - | [130] |

## D  Solutions for SMEs

Table 15: SME Solutions for Data Inventory

| Name / Company | Functionality | Automation | Ref. |
| --- | --- | --- | --- |
| 1Touch | Discovery, Flow | ++ | [3] |
| 2BAdvice | Mapping | | [4] |
| 2BM | Discovery | | [89] |
| Auditrunner | | | [14] |
| BigID | Discovery | | [7] |
| Caralegal | | | [58] |
| Centrl | Monitoring, Mapping | | [145] |
| Clarip | Discovery, Mapping, Flow | | [144] |
| Clym | Mapping | | [149] |
| ContextSpace | Mapping | | [27] |
| Cytrio | Discovery, Mapping | | [37] |
| Data443 | Discovery | | [31] |
| DataGrail | Discovery, Mapping | + | [48] |
| DataGuard | | | [35] |
| Datalegaldrive | Mapping | | [124] |
| DataPrivacyManager | Discovery, Mapping, Flow | | [103] |
| DataReporter | Flow | | [68] |
| DataSentinel | Discovery, Mapping | | [164] |
| Datenschutzverwaltung | Flow | | [30] |
| DPOrganizer | Mapping | | [94] |
| Ecomply | Mapping | | [70] |
| Egnyte | Discovery | | [139] |
| Ethyca | Mapping | ++ | [38] |
| Exterro | Discovery, Mapping | | [72] |
| Heureka | Discovery | | [85] |
| Informatica | Discovery | | [39] |
| ItGovernance | Mapping | | [91] |
| Ketch | Discovery | | [119] |
| Securiti | Discovery, Mapping | + | [123] |
| LightBeam | Discovery | | [157] |
| LogicManager | | | [84] |
| magedata | Discovery, Flow | | [71] |
| Mandatly | Mapping, Flow | | [132] |
| MightyTrust | Mapping, Flow | | [11] |
| Mine | Mapping | | [127] |
| Netwrix | Discovery | | [83] |
| Northdoor | Discovery | | [46] |
| Ohalo | Discovery | | [133] |

+ = at least partially automated — ++ = periodic or continuous automation

| Name / Company | Functionality | Automation | Ref. |
|---|---|---|---|
| OmniPrivacy | | | [188] |
| OneTrust | Discovery, Mapping, Flow | + | [232] |
| opentext | | | [134] |
| OpswareData | Discovery, Mapping | + | [32] |
| Osano | Discovery, Mapping | + | [184] |
| PieEye | Discovery, Mapping | | [136] |
| PiiTools | Discovery | | [87] |
| PKPrivacy | Discovery | | [138] |
| Privacytools | Discovery, Mapping | | [151] |
| Priverion | Flow | | [44] |
| privIQ | Mapping | | [25] |
| Segment | Discovery | | [156] |
| ProteusCyber | Discovery, Mapping | | [226] |
| Cloud Compliance | Discovery | | [24] |
| Securiti | Discovery, Mapping | + | [162] |
| SecuvyAi | Mapping | | [40] |
| Seers | Discovery | | [163] |
| SixFifty | Mapping | | [167] |
| smartprivacy | Mapping | | [243] |
| Sovy | | | [214] |
| Transcend | Discovery, Mapping | | [173] |
| Trunomi | Mapping | | [41] |
| TrustArc | Mapping, Flow | | [176] |
| Truyo | Mapping | | [205] |
| Varonis | Mapping | | [17] |
| VigilantSoftware | Mappping, Flow | | [82] |
| wirewheel | Discovery | | [183] |
| Wrangu | Discovery, Mapping, Flow | | [33] |

+ = at least partially automated — ++ = periodic or continuous automation

Table 16: SME Solutions for data request subjects

| Name / Company | Functionality | Automation | Ref. |
|---|---|---|---|
| SaaS | | | |
| 2BAdvice | | + | [4] |
| 4Comply | Web-Form | + | [5] |
| Adzapier | | + | [9] |
| Akarion | | | [45] |
| Auditrunner | | | [14] |
| BigID | | + | [7] |

+ = at least partially automated — ++ = fully automated

| Name / Company | Functionality | Automation | Ref |
|---|---|---|---|
| Centrl | Forms | | [145] |
| **Add-ons** | | | |
| 2BM | | - | [89] |
| Addison | | | [8] |
| c4b | | | [186] |
| Caralegal | | | [58] |
| Clarip | Web-Form | ++ | [144] |
| Clym | Intake Widget | | [149] |
| Concord | | | [36] |
| ContextSpace | | + | [27] |
| ControlMyID | | + | [28] |
| Corporator | | + | [221] |
| Cytrio | | + | [37] |
| Data443 | | | [31] |
| DataGrail | Web-Form | + | [48] |
| Datagross | | | [49] |
| DataGuard | | | [35] |
| Datalegaldrive | | | [124] |
| DataPrivacyManager | | + | [103] |
| DataReporter | | | [68] |
| DataSentinel | | | [164] |
| Datenschutzverwaltung | | | [30] |
| Defendocs | | | [250] |
| DPOrganizer | | | [94] |
| Juraforum | Response Templates | | [65] |
| EasyGDPR | | + | [69] |
| Ecomply | | | [70] |
| Egnyte | | + | [139] |
| Enactia | | | [73] |
| Ethyca | | + | [38] |
| Exterro | | | [72] |
| Granite | | | [88] |
| GrowthDot | Deletion only | | [81] |
| Heureka | | | [85] |
| IConfirm | | + | [113] |
| Illow | deletion only | + | [116] |
| Informatica | | | [39] |
| Infreemation | SAR only | + | [225] |
| Intellior | | | [52] |
| Isenlabs | Webform | | [90] |
| ItGovernance | | | [91] |
| Ketch | | + | [119] |
| Kiteworks | only deletion? | | [22] |

+ = at least partially automated — ++ = fully automated

| Name / Company | Functionality | Automation | Ref |
|---|---|---|---|
| Securiti | Webforms | + | [123] |
| LightBeam |  | + | [157] |
| LogicGate |  | + | [43] |
| LogicManager | only SAR? |  | [84] |
| magedata |  | + | [71] |
| Mandatly | Webforms | + | [132] |
| Microfocus |  |  | [19] |
| MightyTrust |  |  | [11] |
| Mine |  | + | [127] |
| Netwrix |  | + | [83] |
| Northdoor | SAR only |  | [46] |
| Ohalo | SAR only? |  | [133] |
| Omikron |  | + | [131] |
| OmniPrivacy |  |  | [188] |
| OneTrust | Web-forms | + | [232] |
| opentext |  |  | [134] |
| OpswareData | SAR only?, Webforms | + | [32] |
| Osano | Web-Forms | + | [184] |
| PieEye |  |  | [136] |
| PiiTools |  |  | [87] |
| PiwikPRO | Web-Forms |  | [137] |
| PKPrivacy |  | + | [138] |
| Prighter | Workflows | + | [47] |
| Privacy1 |  | + | [140] |
| PrivacyBunker | Self-Service |  | [248] |
| PrivacyEngine | SAR only? |  | [234] |
| PrivacyNexus |  |  | [146] |
| PrivacyPerfect | Web-Form | + | [235] |
| Privacytools |  | + | [151] |
| Priverion |  |  | [44] |
| privIQ |  |  | [25] |
| Segment |  | + | [156] |
| ProteusCyber |  |  | [226] |
| RealDPG |  |  | [158] |
| Responsum |  | + | [195] |
| Cloud Compliance |  | + | [24] |
| Securiti |  | + | [162] |
| SecuvyAi | Web-Form | + | [40] |
| Seers |  |  | [163] |
| SixFifty |  |  | [167] |
| smartprivacy |  |  | [243] |
| SureCloud | SAR only? | + | [96] |
| Transcend | WebForm, API | ++ | [173] |

+ = at least partially automated — ++ = fully automated

| Name / Company | Functionality | Automation | Ref |
|---|---|---|---|
| Trunomi | | + | [41] |
| TrustArcLeaderPrivacy | | + | [176] |
| Truyo | | + | [205] |
| Usoft | | | [187] |
| Varonis | SAR only?, Webforms | | [17] |
| VigilantSoftware | SAR only? | | [82] |
| WeControl | SAR only? | | [180] |
| wirewheel | SAR only? | + | [183] |
| DSGVO All In One | Webforms | | [224] |
| Trew Knowledge | | | [217] |
| audrasjb | Webform | | [190] |
| Wrangu | | + | [33] |

\+ = at least partially automated — ++ = fully automated

Table 17: SME Solutions for consent management

| Name / Company | Type | Ref. |
|---|---|---|
| 4Comply | U | [5] |
| Adzapier | C | [9] |
| c4b | U | [186] |
| Centrl | UC | [145] |
| Clarip | UC | [144] |
| Clym | C | [149] |
| Complianz | C | [150] |
| Concord | UC | [36] |
| ConsentGrid | U | [220] |
| CookieScript | C | [29] |
| Data443 | U | [31] |
| DataGuard | UC | [35] |
| DataPrivacyManager | U | [103] |
| DataReporter | U | [68] |
| Dataships | C | [42] |
| Defendocs | C | [250] |
| Egnyte | U | [139] |
| Exterro | UC | [72] |
| IConfirm | U | [113] |
| Illow | C | [116] |
| Isenlabs | C | [90] |
| Iubenda | C | [86] |
| Ketch | UC | [119] |
| Securiti | UC | [123] |

| Name / Company | Type | Link |
|---|---|---|
| Legalweb | C | [125] |
| LightBeam | UC | [157] |
| Mandatly | C | [132] |
| Mine | UC | [127] |
| OneTrust | UC | [232] |
| opentext |  | [134] |
| Osano | C | [184] |
| PiwikPRO | UC | [137] |
| Privacy1 | C | [140] |
| PrivacyBunker | C | [248] |
| PrivacyPerfect | UC | [235] |
| PrivacyPolicies | C | [67] |
| Privacytools | UC | [151] |
| Segment | U | [156] |
| ProteusCyber | UC | [226] |
| Responsum | C | [195] |
| Cloud Compliance |  | [24] |
| SecurePrivacy | C | [34] |
| Securiti | UC | [162] |
| SecuvyAi | U | [40] |
| Seers | U | [163] |
| SimplyGDPR |  | [166] |
| SixFifty | C | [167] |
| Sovy | C | [214] |
| Termly | C | [101] |
| Transcend | C | [173] |
| Truendo | C | [175] |
| Trunomi | UC | [41] |
| Transcend | UC | [176] |
| Truyo | C | [205] |
| UserCentrics | C | [168] |
| wirewheel | U | [183] |
| DSGVO All In One | C | [224] |
| Trew Knowledge |  | [217] |
| Wrangu |  | [33] |

Table 18: SME Solutions for data notices

| Name / Company | Functionality | Ref. |
|---|---|---|
| Clym | Management | [149] |
| Complianz | Generaror | [150] |
| CookieScript | Generator | [29] |
| Cortina Consult | Generator, Management | [53] |

| Name / Company | Functionality | Ref. |
| --- | --- | --- |
| DataGuard | | [35] |
| Datalegaldrive | Templates | [124] |
| DataReporter | | [68] |
| DataSentinel | | [164] |
| Dataships | | [42] |
| Datenschutz.org | Template | [54] |
| Weiß & Partner | Generator | [56] |
| datenschutz-generator | Generator | [55] |
| Defendocs | Generator | [250] |
| EasyGDPR | | [69] |
| Ecomply | Templates | [70] |
| e-recht24 | Generator | [241] |
| GetTerms | Generator | [179] |
| Haendlerbund | Generator | [159] |
| Illow | | [116] |
| Intent | Generator | [66] |
| Iubenda | Generator | [86] |
| Securiti | Templates, Management | [123] |
| Legalweb | | [125] |
| MightyTrust | | [11] |
| OneTrust | Templates, Management | [232] |
| Privacy1 | | [140] |
| PrivacyPolicies | Generator | [67] |
| PrivacyPolicyOnline | Generator | [148] |
| Privacytools | | [151] |
| Priverion | Templates | [44] |
| Cloud Compliance | | [24] |
| SecurePrivacy | Generator | [34] |
| Securiti | Templates, Management | [162] |
| Seers | | [163] |
| Shopify Privacy Notice Generator | Generator | [147] |
| SimplyGDPR | Generator | [166] |
| smartprivacy | | [243] |
| Sovy | Generator | [214] |
| Termly | Generator | [101] |
| TermsFeed | Generator | [171] |
| Truendo | Generator | [175] |
| Truyo | | [205] |
| WBSDatenschutzgenerator | Generator | [57] |
| WeControl | | [180] |
| wirewheel | | [183] |
| DSGVO All In One | Generator | [224] |
| Trew Knowledge | | [217] |
| Wrangu | | [33] |

Table 19. SME Solutions for data subject portals

| Name / Company | Functionality | Ref. |
| --- | --- | --- |
| 4Comply | | [5] |
| Clarip | | [144] |
| Cytrio | | [37] |
| Data443 | | [31] |
| DataGuard | | [35] |
| Datalegaldrive | | [124] |
| DataPrivacyManager | | [103] |
| DataSentinel | | [164] |
| Dataships | | [42] |
| DPOrganizer | | [94] |
| Ethyca | | [38] |
| Exterro | | [72] |
| IConfirm | | [113] |
| Securiti | DSR, Notice, Consent | [123] |
| LogicGate | | [43] |
| Mine | | [127] |
| OmniPrivacy | | [188] |
| OneTrust | DSR, Notice, Consent | [232] |
| Priverion | | [44] |
| Segment | | [156] |
| ProteusCyber | | [226] |
| Securiti | | [162] |
| Transcend | | [173] |
| Trunomi | | [41] |
| Truyo | | [205] |

Table 20. Solutions small and medium enterprises for data breach handling

| Name / Company | Functionality | Ref. |
| --- | --- | --- |
| 1Touch | | [3] |
| Akarion | | [45] |
| Auditrunner | | [14] |
| Caralegal | | [58] |
| Datalegaldrive | | [124] |
| DataSentinel | Notification | [164] |
| DPOrganizer | | [94] |
| Ecomply | | [70] |
| Egnyte | | [139] |
| Enactia | | [73] |
| Granite | | [88] |
| IConfirm | | [113] |
| Intellior | | [52] |
| ItGovernance | | [91] |
| Securiti | Management, Notification | [123] |
| LightBeam | | [157] |
| LogicGate | | [43] |
| LogicManager | Notification | [84] |
| MightyTrust | | [11] |
| Northdoor | | [46] |
| OmniPrivacy | | [188] |
| OneTrust | Management, Notification | [232] |
| opentext | | [134] |
| PiiTools | | [87] |
| PKPrivacy | | [138] |
| PrivacyNexus | | [146] |
| PrivacyPerfect | | [235] |
| Privacytools | | [151] |
| Priverion | | [44] |
| privIQ | | [25] |
| ProteusCyber | | [226] |
| RealDPG | | [158] |
| Responsum | | [195] |
| Seers | | [163] |
| SureCloud | | [96] |
| Trunomi | | [41] |
| Usoft | | [187] |
| VigilantSoftware | | [82] |
| wirewheel | | [183] |
| Trew Knowledge | | [217] |
| Wrangu | | [33] |

Table 21. Part1: comparison of example solutions for small and medium enterprises

| Comparison | OneTrust | Osano | Transcend | Datagrail | Securiti |
|---|---|---|---|---|---|
| Data Discovery | ✓ | ✓ | ✓ | ✓ | ✓ |
| Data Mapping | ✓ | ✓ | ✓ | ✓ | ✓ |
| Connect joint controllers | ? | ? | ? | ? | ? |
| Automation | ✓ | ✓ | ✓ | ✓ | ✓ |
| Add custom stores | ✓ | x | ✓ | ✓ | ? |
| Pre-Built Connectors[a] | 61 for DD[b] + more | 132 | 1300+ | 1400+ | 1000+ |
| RoPA [c] | dynamic | provides reference | dynamic | dynamic | dynamic |
| DSR types | All | All | All | All | Access, Erasure, ? |
| Request channel | Form | Form, Email | Form, API, Other | Form | Form |
| Data transmission channel | Portal | Portal | Portal | Link via mail | Portal |
| DSAR Data types | ? | ? | ? | ? | ? |
| Data presentation | ? | ? | ? | ? | ? |
| Authentication types[d] | 2FA, SSO/OIDC, 3rd-party, ID | any file | 2FA, JWT, OAuth2 | 2FA, Questions[e] | Unspecified |
| Authentication channel | ? | Form, Portal | ? | ? | Portal |
| Data Import | ? | ? | ? | ? | ? |
| Possible Automation | Full | Partial | Full | Partial, maybe full | Partial, maybe full |
| Data Breach management | ✓ | x | x | x | ✓ |
| Data Breach notice | ✓ | x | x | x | ✓ |

[a] Transcend, Datagrail and Securiti don't offer filtering and counting of DD integration
[b] Data Discovery
[c] dynamic means, that the RoPA is at least partially populated and updated automatically by the system. Manual intervention and additions may be required
[d] 2FA by email and/or SMS
[e] Questions about existing data

Table 22. Part2: comparison of example solutions for small and medium enterprises

| Comparison | OneTrust | Osano | Transcend | Datagrail | Securiti |
| --- | --- | --- | --- | --- | --- |
| Cookie Consent | ✓ | ✓ | ✓ | x | ✓ |
| Custom Forms | ✓ (+ builder) | x | x | x | ✓ (+ builder) |
| Universal Consent | ✓ | x | x | x | ✓ |
| Mobile Apps | ✓ | ✓[a] | - | x | x |
| Cookie Scan | ? | Initial | Continuously | x | Periodic |
| Classification | ? | Automatic | Automatic | x | Automatic |
| Explains obscure cookies | Some basic, some detailed | x | Basic | x | Some basic, some detailed |
| DnT / DnS / GPC | ? | ? / ✓ / ✓ | ✓ / ✓ / ✓ | x | ? / ? / ✓ |
| Consent Record | ? | ✓ | x | x | ✓ |
| Data Templates | ✓ | ✓ | x | x | ✓ |
| Data Management | ✓ | x | x | x | ✓ |
| Data Versioning | ✓ | x | x | x | x |
| Scan for policies | Regularly | x | x | x | x |
| Dynamic population | Cookies | x | Data Practices | x | Cookies, Consent, Processing, DSR |
| Visual help | Structured, Overview | x | x | x | x |
| Privacy Center | Notices, DSR, consent | x | Notices, DSR, consent ?[b] | | Notices, DSR, consent[c] |
| Training | ✓ | ✓ | x | x | ✓ |
| Offers demo | ✓ | ✓ | ✓ | ✓ | ✓ |

[a] Beta
[b] Osano has one on their website but don't mention one as product
[c] Securiti's privacy center is early access